\documentclass[journal=ancac3,manuscript=article]{achemso}

\usepackage[version=3]{mhchem}
\usepackage{bm}
\usepackage{amsmath}
\usepackage{amssymb}
\renewcommand{\vec}[1]{\bm{#1}}

\makeatletter
\renewcommand{\thefigure}{S\@arabic\c@figure}

\makeatother

\author{S{\'{e}}rgio L. Morelh{\~{a}}o}
\email{morelhao@if.usp.br}
\affiliation{Department of Physics, University of Guelph, Guelph, Ontario
N1G\,1W2, Canada}
\alsoaffiliation{Institute of Physics, University of S\~ao Paulo,
S\~ao Paulo 05508-090, Brazil}

\author{Stefan Kycia}
\affiliation{Department of Physics, University of Guelph,
Guelph, Ontario N1G\,1W2, Canada}
\email{skycia@uoguelph.ca}

\author{Samuel Netzke}
\affiliation{Department of Physics, University of Guelph, Guelph, Ontario N1G\,1W2, Canada}

\author{Celso I. Fornari}
\affiliation{National Institute for Space
Research, S\~ao Jos\'e dos Campos, S\~ao Paulo 12227-010, Brazil}

\author{Paulo H. O. Rappl}
\affiliation{National Institute for Space
Research, S\~ao Jos\'e dos Campos, S\~ao Paulo 12227-010, Brazil}

\author{Eduardo Abramof} %\email[]{eduardo.abramof@inpe.br}
\affiliation{National Institute for
Space Research, S\~ao Jos\'e dos Campos, S\~ao Paulo 12227-010, Brazil}

\title{Lateral lattice coherence lengths in thin films of bismuth telluride topological insulators, with overview on polarization factors for X-ray dynamical diffraction in monochromator crystals}

\begin{document}

\begin{abstract}
In the supporting information file for article Dynamics of Defects in van der Waals Epitaxy of Bismuth Telluride Topological Insulators (J. Phys. Chem. C 2019, 123, 24818-24825, doi: 10.1021/acs.jpcc.9b05377), several topics on X-ray diffraction analysis of thin films were developed or revisited. A simple equation to determine lateral lattice coherence lengths in thin films stands as the main development (section S4 - Lateral lattice coherence length in thin films), while X-ray dynamical diffraction simulation in monochromator crystals stands as an interesting overview on how the ratio between $\pi$ and $\sigma$ polarization components is affected by whether diffraction takes place under kinematical or dynamical regime (section S3 - Polarization factor).
\end{abstract}

\section{S1 - Choice of asymmetric reflections}
All allowed hkl reflections measured in a recent work\cite{sm19} are listed in Table~\ref{tab:reflections}. Forbidden reflections with khl indexes, h$\rightleftarrows$k regarding those in the Table~\ref{tab:reflections}, have null structure factors. They were also measured due to twinned domains in the films that are 60$^\circ$ rotated in azimuth regarding the film main lattice \cite{cf16a,cf16b}.

\begin{table}[ht]
 \renewcommand\thetable{S1}
\scriptsize{
\caption{List of allowed reflections measured in a recent work\cite{sm19}. Diffraction vector of modulus $Q=(4\pi/\lambda)\sin\theta$, instrumental angles $\theta$ (detector arm at $2\theta$), $\chi$, and $\phi$, incidence angle $\alpha_{\rm i}$, and polarization factor $p$ from Eq.~(\ref{eq:p2th}). Structure factors $F$ calculated for the Bi$_2$Te$_3$ crystal structure (lattice parameters $a=4.382$\AA\, and $c=30.497$\AA)\cite{hs14} using resonant amplitudes \cite{sm16,fpfpp} and null Debye-Waller factors. ${\rm h}_s{\rm k}_s{\rm l}_s$ stand for film reflection indexes regarding the substrate reciprocal lattice (only for the 1st reflections in column 1).}\label{tab:reflections}
\begin{tabular}{ccccccccccc}
  \hline\hline
  hkl & $Q$\,(\AA$^{-1}$) & $\theta$\,($^\circ$) & $\chi$\,($^\circ$) & $\phi$\,($^\circ$) & $\alpha_{\rm i}$\,($^\circ$) & $p$ & $|F|$ & h$_s$ & k$_s$ & l$_s$\\
  \hline
 $01\,5/\bar{1}0\,5/1\bar{1}\,5$ & 1.950 & 13.831 & 31.889 & -60/60/180 & 7.26 & 0.679 & 815.0 &  1.25 & -0.75 &  1.25 \\
 $10\,10/\bar{1}1\,10/0\bar{1}\,10$ & 2.643 & 18.907 & 51.214 & -120/0/120 & 14.63 & 0.642 & 725.5 & 0.51 &  0.51 &  2.51\\
 $20\,5/\bar{2}2\,5/0\bar{2}\,5$ & 3.468 & 25.160 & 17.280 & -120/0/120 & 7.26 & 0.593 & 695.5 &-0.75 & -0.75 &  3.25\\
 $02\,10/\bar{2}0\,10/2\bar{2}\,10$ & 3.900 & 28.562 & 31.889 & -60/60/180 & 14.63 & 0.567 & 637.6 &  2.51 & -1.49 &  2.51\\
 $\bar{2}\bar{1}\,5/3\bar{2}\,5/\bar{1}3\,5$ & 4.500 & 33.482 & 13.233 & 79/199/319 & 7.26 & 0.535 & 626.1 & 1.25 & 3.25 & -2.75\\
 $21\,10/\bar{3}2\,10/1\bar{3}\,10$ & 4.841 & 36.403 & 25.189 & -101/19/139 & 14.63 & 0.520 & 579.7 &  0.51 & -1.49 &  4.51 \\
  \hline
  \hline
\end{tabular}
}
\end{table}

For thin epitaxial films undergoing Kinematical diffraction, the integrated intensity of a Bragg reflection is proportional to the beam footprint on the film surface, Eq.~(\ref{eq:Phkl}). Then, to improve accuracy in determining atomic displacement parameters from integrated intensity data, the preference is for sets of reflections that have a common incidence angle $\alpha_{\rm i}$. With the film surface normal direction $\hat{\vec{n}}$, set collinear to the $\phi$ rotation axis of the 4-circle goniometer (Fig. 2a in the main text), the incidence angle can be obtained from the goniometer angles as $\sin\alpha_{\rm i}=\sin\theta\,\sin\chi$. When the diffraction vector $\vec{Q}$ of an asymmetric reflection is placed in the incidence plane at the correct Bragg angle, $\sin\theta = (\lambda/4\pi)Q$ and $\sin\chi=\vec{Q}\cdot\hat{\vec{n}}/Q$, leading to
\begin{equation}\label{eq:alphai}
    \sin\alpha_{\rm i} = (\lambda/4\pi)\vec{Q}\cdot\hat{\vec{n}} = \lambda {\rm l}/2c\,,
\end{equation}
which is constant for all asymmetric reflections with the same l index in (001) films. Two sets of asymmetric reflections were chosen, hk5 and hk10 with incidence angles $\alpha_{\rm i}=7.26^\circ$ and $14.63^\circ$, respectively.

\begin{figure}
  \includegraphics[width=6.2in]{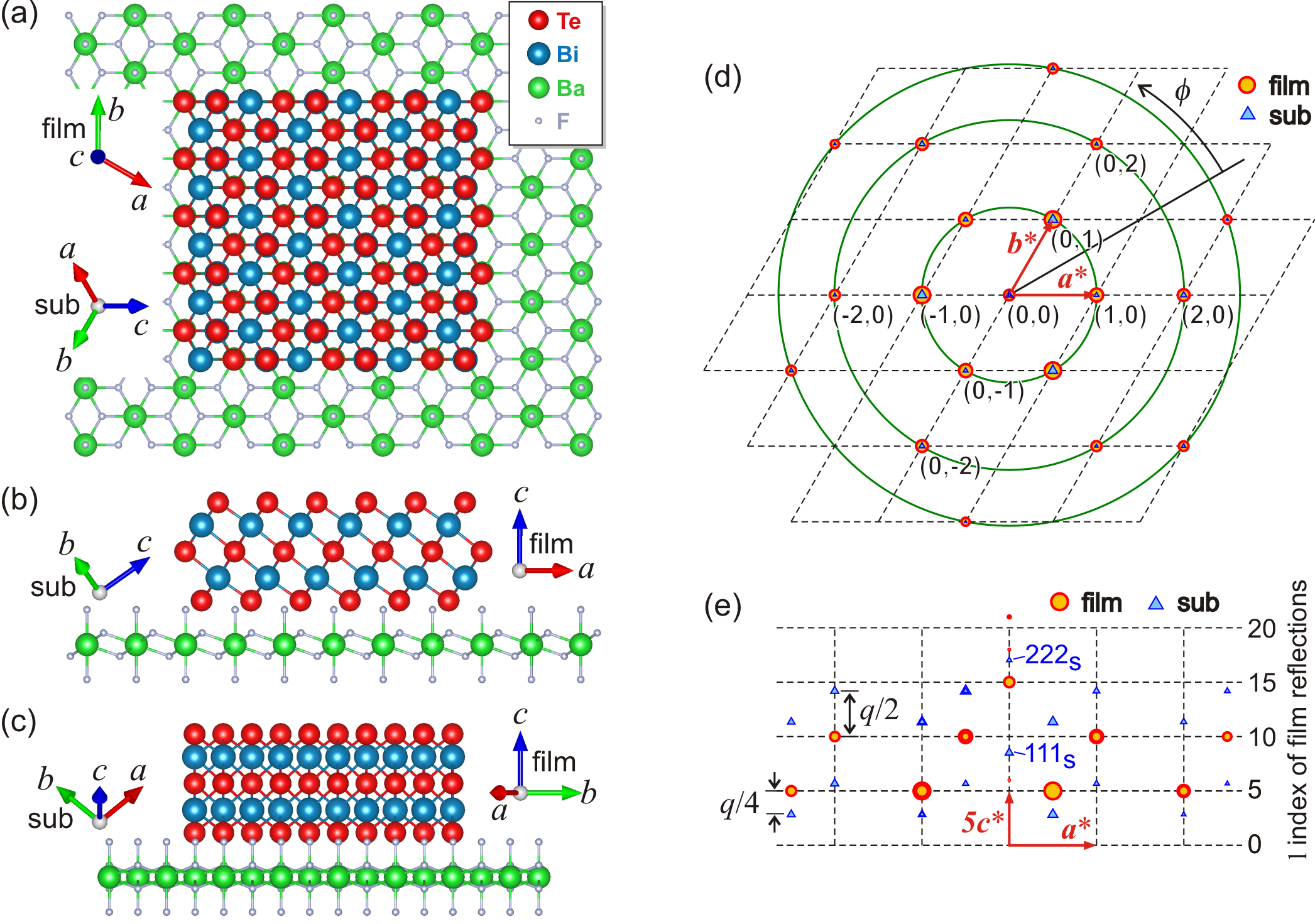}\\
  \caption{(a-c) Atomic arrangement of hexagonal Bi$_2$Te$_3$ film on cubic BaF$_2$ (111) substrate seen from different perspectives, as indicated by arrows along the basis vectors $\vec{a}$, $\vec{b}$, and $\vec{c}$ of each lattice. (d,e) Film and substrate reciprocal lattice points (RLPs) according to the lattices relative orientation given in the left panels where $[110] {\rm Bi}_2{\rm Te}_3 (001)\,||\,[0\bar{1}1] {\rm BiF}_2(111)$, azimuth $\phi=0$.  $\vec{a}^*$, $\vec{b}^*$, and $\vec{c}^*$ stand for basis vectors of the film reciprocal lattice only. (d) In-plane viewing with RLPs connected by circles of radius $|{\rm h}\vec{a}^*+{\rm k}\vec{b}^*|=1.654$\,\AA$^{-1}$, 3.308\,\AA$^{-1}$, and 4.376\,\AA$^{-1}$. (e) Film RLPs (orange circles) of the measured asymmetric reflections falling either at 1/4 or 1/2 of the distance $q=2\pi/d_{111_S}=1.755\,\textrm{\AA}^{-1}$ between adjacent substrate RLPs (blue triangles) aligned along the surface normal direction. }\label{fig:recspaceviews}
\end{figure}

Besides reflections with a common angle of incidence, film reflections must be far way from the substrate reflections to avoid extra intensity contributions in the film rocking curves ($\theta$-scans) for integrated intensity measurements. The relative orientation of film and substrate lattices, as depicted in Fig.~\ref{fig:recspaceviews}, is such that all allowed hk5 and hk10 film reflections are aligned along the surface normal direction with substrate reflections. However, their reciprocal lattice points (RLPs) fall in between those from the substrate lattice at a minimum distance of $\Delta Q_z = 0.44\,\textrm{\AA}^{-1}$ in reciprocal space, Fig.~\ref{fig:recspaceviews}(e). The closest substrate diffracted beam propagates at $\Delta\chi\approx6^\circ$ from the incidence plane, as given by $$\Delta \chi \simeq \vec{Q}\cdot\hat{\vec{n}}/Q - (\vec{Q}\cdot\hat{\vec{n}}-\Delta Q_z)/\sqrt{Q^2+\Delta Q_z^2\,}$$ for $\vec{Q} = \bar{2}\vec{a}^* +  \bar{1}\vec{b}^* +  5\vec{c}^*$ ($Q = 4.496\,\textrm{\AA}^{-1}$), which is easily cut off by the vertical acceptance of $0.6^\circ$ of the detector system.

Choosing only reflections that have three-fold symmetry around the [001] direction is also important due to twinned domains often observed in the Bi$_2$Te$_3$ films. Since these domains are rotated by $60^\circ$ in azimuth regarding the main lattice of the films, reflections with six-fold symmetry are inevitably
 mixing contributions from the main lattice and from twinned domains. One the other hand, allowed reflections such as $01\,5$, $\bar{1}0\,5$, and $1\bar{1}\,5$ are set apart by $120^\circ$ in azimuth, as given in Table~\ref{tab:reflections}, while reflections $10\,5$, $0\bar{1}\,5$, and $\bar{1}1\,5$ are forbidden, unless receiving intensity contributions from twinned domains.

\section{S2 - Integrated intensity}

In small single crystals and thin epitaxial films, atomic displacement values can be determined by measuring the diffraction power $P=\int I(\theta){\rm d}\theta =I_e|F|^2 N \lambda^3/\sin(2\theta)\, V_{\rm cel}$\, of different hkl reflections \cite{sm16}; it is also known as integrated intensity of the diffraction curve $I(\theta)$ as a function of the rocking curve angle $\theta$. For thin films, this general expression can be written in terms of three parameters that are varying from one reflection to another: the scattering angle $2\theta$, the structure factor $F$, and the number $N$ of unit cells within the diffracting volume $NV_{\rm cell}$ for x-ray of wavelength $\lambda$. The scattering intensity by a single electron, $I_e$, also depends on $2\theta$ through the polarization factor $p$ since $I_e\,\propto\,p$. In thin films of uniform thickness and negligible absorption, $N$ is proportional to the beam footprint $S_0/\sin\alpha_{\rm i}$ for an incident x-ray beam of cross-section $S_0$, leading to
\begin{equation}\label{eq:Phkl}
    P=C\,p\,|F|^2 /(\sin2\theta\sin\alpha_{\rm i})
\end{equation}
where $C$ is a constant for each sample. Since the number of accessible reflections in thin films are limited, the atomic displacement parameters $U_{\rm ij}$,\cite{wk03} for all elements have been restricted to the diagonal terms only, $U_{11}= U_{22}=U_y$ and $U_{33} = U_z$ with respect to in-plane (or lateral) and longitudinal directions, respectively. With this restriction the structure factor expression simplifies to $F = {\rm exp}[-\frac{1}{2}(Q_{y}^2 U_{y}+Q_{z}^2 U_{z})] \sum_a f_a {\rm exp}(i\vec{Q}\cdot\vec{r}_a)$
where the diffraction vector $\vec{Q} = {\rm h}\vec{a}^* + {\rm k}\vec{b}^* + {\rm l}\vec{c}^*$ has been splitted into two components: $\vec{Q}_{y} = {\rm h}\vec{a}^* + {\rm k}\vec{b}^*$ in the plane of the film, Figure.~\ref{fig:recspaceviews}(d), and $\vec{Q}_z = {\rm l}\vec{c}^*$ along the growth direction. The corresponding root mean square (rms) atomic displacements are then $U_y^{1/2}$ (lateral) and $U_z^{1/2}$ (longitudinal). Data fitting of experimental integrated intensities (peak areas) were carried out by using a simulated annealing (SA) algorithm \cite{kir83,kab17} to adjust $C$, $U_y$, and $U_z$ in the above equation, Eq.~(\ref{eq:Phkl}). These fitting parameters were adjusted by minimizing the mean squared logarithmic error (MSLE) $E=\frac{1}{N_j}\sum_j{\ln^2{[P_j/P_s(\vec{Q}_j, C, U_y, U_z)}]}$ where $P_j$ and $P_s (\vec{Q}_j,C,U_y,U_z)$ are experimental and simulated data points for each j-th reflection of diffraction vector $\vec{Q}_j$ in the set of $N_j$ reflections.  After minimizing the MSLE function, relative variation of the experimental data due to atomic displacements have been displayed in the main text as
\begin{equation}\label{eq:DP_P}
    \Delta P/P = \frac{P_j - P_s(\vec{Q}_j, C, 0, 0)}{P_s(\vec{Q}_j, C, 0,0)}\,.
\end{equation}

\section{S3 - Polarization factor}

\begin{figure}[t]
  \includegraphics[width=5.4in]{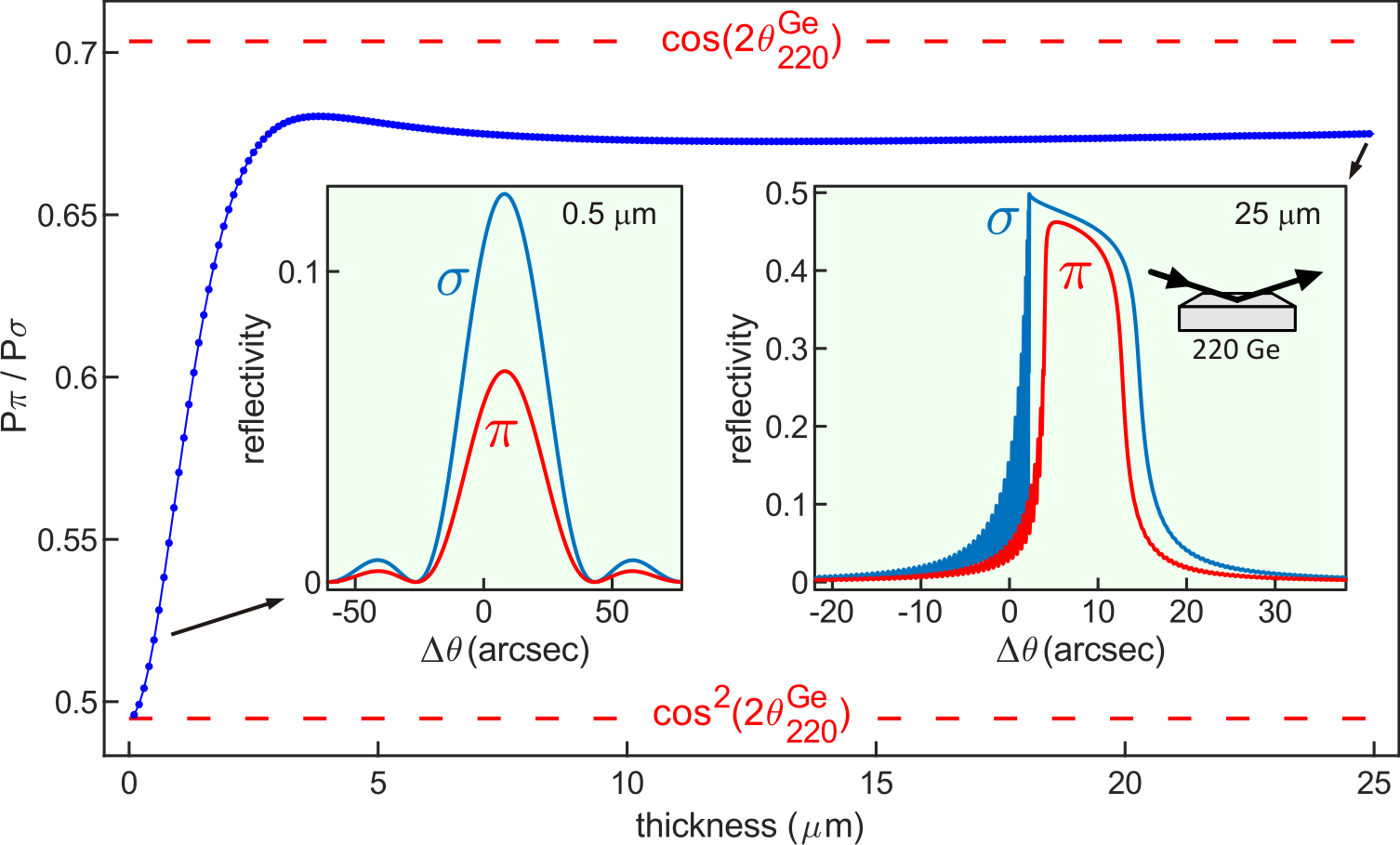}\\
  \caption{For a single 220 Ge Bragg reflection, ratio $R_{\pi\sigma}$ between integrated reflectivities $P_\pi$ and $P_\sigma$ of each polarization component in the diffracted beam as a function of crystal thickness. Cu$K_{\alpha1}$ radiation ($\theta^{\rm Ge}_{220}=22.6484^\circ$) with $45^\circ$ linearly polarized incident beam, i.e. $\hat{\vec{\varepsilon}}=\cos(45^\circ)\hat{\vec{\pi}} + \sin(45^\circ)\hat{\vec{\sigma}}$ in Eq.~(\ref{eq:P}). Examples of reflectivity curves from dynamic diffraction simulation\cite{wec97} are given in the insets. For crystals undergoing kinematic diffraction $R_{\pi\sigma}=\cos^2(2\theta^{\rm Ge}_{220})$, while under dynamical regime of diffraction $R_{\pi\sigma}\lesssim\cos(2\theta^{\rm Ge}_{220})$.}
  \label{fig:dynamicpolfactor}
\end{figure}

Unpolarized x-rays of wavevector $\vec{K}=(2\pi/\lambda)\hat{\vec{s}}$, after scattering by electrons into wavevector $\vec{K}^\prime=(2\pi/\lambda)\hat{\vec{s}}^\prime$ have polarization factor $p(2\theta)=\langle|\vec{\mathcal{P}}|^2\rangle$ where
\begin{equation}\label{eq:P}
    \vec{\mathcal{P}} = \hat{\vec{s}}^{\prime} \times (\hat{\vec{\varepsilon}} \times \hat{\vec{s}}^{\prime})
\end{equation}
stands for each linearly polarized component of the incident wavefield vibrating along direction $\hat{\vec{\varepsilon}}=\cos(\varepsilon)\hat{\vec{\pi}} + \sin(\varepsilon)\hat{\vec{\sigma}}$ \cite{sm01,sm02c,sm11,sm16}. The two orthogonal components have been defined as $\hat{\vec{\pi}}=\hat{\vec{\sigma}}\times\hat{\vec{s}}$ and $\hat{\vec{\sigma}}=\hat{\vec{s}}\times\hat{\vec{s}}^{\prime}/\sin(2\theta)$. By using $\hat{\vec{s}}=\hat{\vec{z}}$ and  $\hat{\vec{s}}^\prime=\sin(2\theta)\hat{\vec{x}}+\cos(2\theta)\hat{\vec{z}}$, we have that $\vec{\mathcal{P}}=\cos(\varepsilon)\cos^2(2\theta)\hat{\vec{x}} + \sin(\varepsilon)\hat{\vec{y}} - \cos(\varepsilon)\cos(2\theta)\sin(2\theta)\hat{\vec{z}}$. Since $\langle \cos^2(\varepsilon)\rangle=\langle \sin^2(\varepsilon)\rangle=1/2$ for $\varepsilon$ varying from 0 to $2\pi$ in the unpolarized beam, $p(2\theta)=[1+\cos^2(2\theta)]/2$ is the well known polarization factor for scattering of unpolarized x-rays. It is also the polarization factor in the case of x-ray diffraction in small crystals such as thin epitaxial films diffracting according to the Kinematical theory. In large crystals such as the monochromator crystals undergoing a single Bragg reflection in reflection geometry, the intensity ratio $R_{\pi\sigma}$ between the $\pi$ and $\sigma$ components in the diffracted beam is affected by crystalline perfection and x-ray absorption that can be different for each of these components.\cite{aa03} The exact polarization factor for a perfect Ge 220 monochromator and Cu$K_{\alpha 1}$ radiation, $\theta^{\rm Ge}_{220}=22.6484^\circ$, can be obtained by dynamical diffraction simulation,\cite{wec97} as shown in Fig.~\ref{fig:dynamicpolfactor}. For very thin crystals ($< 0.1\,\mu$m) or crystals with damaged surface diffracting kinematically, $R_{\pi\sigma}\simeq\cos^2(2\theta^{\rm Ge}_{220})=0.495$, while for perfect thick crystals ($>5\,\mu$m), $R_{\pi\sigma}=0.675\,<\cos(2\theta^{\rm Ge}_{220})$.

\begin{figure}[ht]
  \includegraphics[width=5.4in]{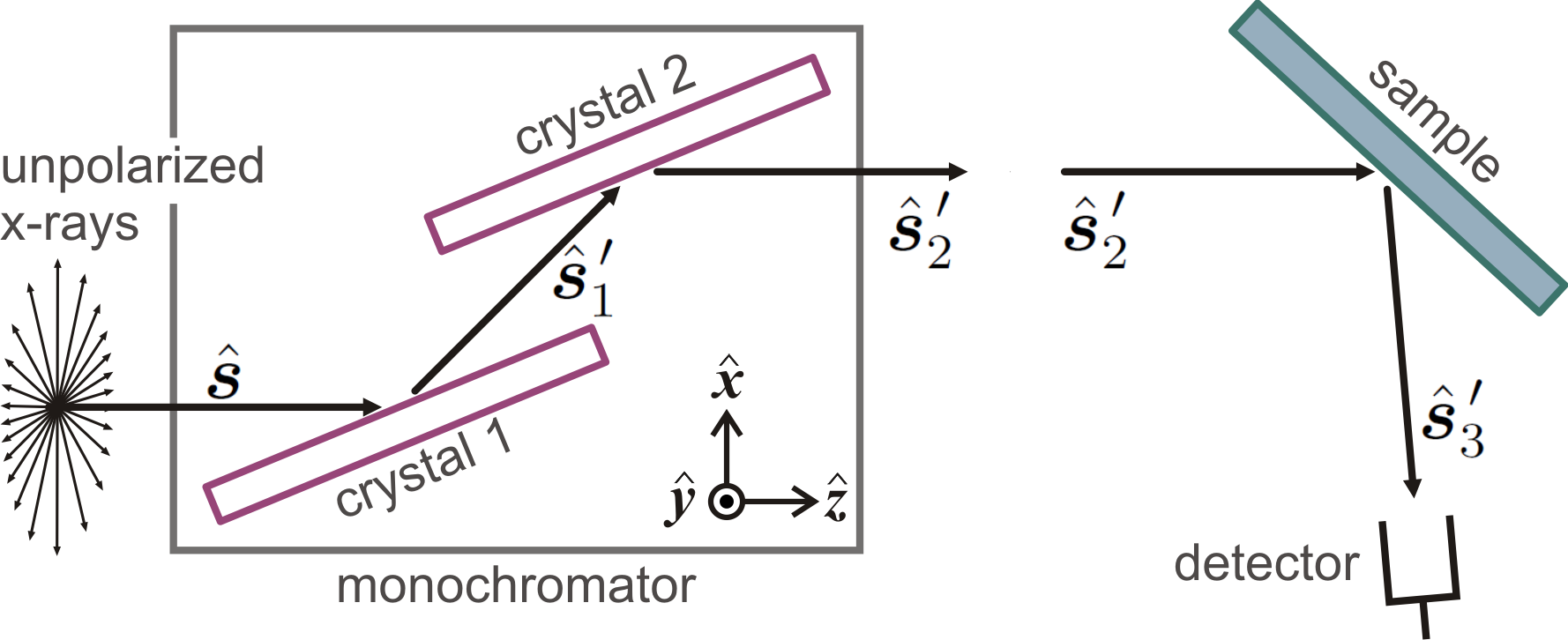}\\
  \caption{Bragg reflections along the beam path from unpolarized x-rays before the monochromator until detection after the sample. Each reflection changes the beam direction to $\hat{\vec{s}}_n^\prime$ and polarization according to Eq.~(\ref{eq:Pn}), leading to the polarization factor $p(2\theta_{\rm hkl})$ in Eq.~(\ref{eq:p2th}).}\label{fig:polfactor}
\end{figure}

\begin{figure}
  \includegraphics[width=3.2in]{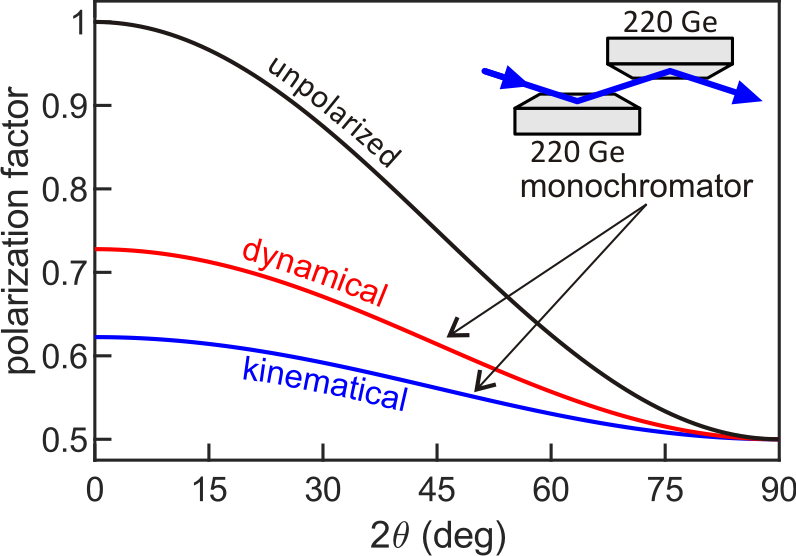}\\
  \caption{Polarization factor as a function of the diffracted beam angle $2\theta$ in the film reflections. Calculation for a two-reflection monochromator (inset) when using $R_{\pi\sigma}^2=0.46$ (dynamical) or 0.25 (kinematical) in Eq.~(\ref{eq:p2th}). For sake of comparison, the polarization factor for unpolarized radiation, i.e. without the monochromator ($R_{\pi\sigma}^2=1$), is also shown.}
  \label{fig:pfactorcomparison}
\end{figure}

After the double collimating multilayer optic of the used diffractometer, the x-ray beam is still unpolarized before hitting the monochromator. In total, the detected x-rays undergo three Bragg reflections, two inside the monochromator and one in the sample, as depicted in Fig.~\ref{fig:polfactor}. Then, the above equation, Eq.~(\ref{eq:P}), has to be applied recursively to each reflection, resulting in
\begin{equation}\label{eq:Pn}
     \vec{\mathcal{P}}_n = \hat{\vec{s}}_n^{\,\prime} \times (\vec{\mathcal{P}}_{n-1} \times \hat{\vec{s}}_n^{\,\prime})
\end{equation}
where $\vec{\mathcal{P}}_0=\hat{\vec{\varepsilon}}$ and $n=1, 2, 3$. By taking $R_{\pi\sigma}$ as the reduction ratio in the $\pi$ component after each 220 Ge reflection inside the monochromator, and $\theta_{\rm hkl}$ as the Bragg angle of reflection hkl in the film,
\begin{equation}\label{eq:p2th}
    p(2\theta_{\rm hkl}) = \left [ 1+R_{\pi\sigma}^2\cos^2(2\theta_{\rm hkl})\right ]/2
\end{equation}
is the final polarization factor to be used when calculating the integrated intensities of the film's hkl reflections. It implies that, the monochromator delivers x-rays with a relative amount $R_{\pi\sigma}^2$ of polarization in the incidence plane of the diffractometer. This component of $\pi$-polarization is the fraction of incident x-rays in the sample that are in fact susceptible to the diffraction angle $2\theta_{\rm hkl}$ of the film reflections. Without accounting for polarization in the monochromator ($R_{\pi\sigma}^2=1$), there would be a much more drastic reduction in the relative values of integrated intensities as the diffraction angle increases, as shown in Fig.~\ref{fig:pfactorcomparison}. With the two-reflection monochromator, the relative amount of $\pi$-polarization is in the range $0.25\leq R_{\pi\sigma}^2 \lesssim 0.46$ depending on the diffraction regime (kinematical or dynamical) of the monochromator crystals.  The in-plane rms atomic displacement values reported in a recent work\cite{sm19} were determined for $R_{\pi\sigma}^2 = 0.46$ and the polarization factors listed in Table~\ref{tab:reflections}. By using $R_{\pi\sigma}^2 = 0.25$ instead (kinematical approach), all values are evenly increased by about 2\,pm within the same error bars, i.e. $U_y^{1/2}=$ 14.8\,pm, 14.1\,pm, 13.9\,pm, and 14.1\,pm would be 16.6\,pm, 16.0\,pm, 15.8\,pm, and 15.9\,pm, respectively.

\section{S4 - Lateral lattice coherence length in thin films}

Intensity distribution around reciprocal lattice points (RLPs) are related by Fourier transform to lattice coherence lengths inside the diffracting volume. In a perfect crystal domain, the coherence lengths are the sizes of the domain itself. But in epitaxial films, elastic strain and defects due to accommodation of lattice misfit at the film-substrate interface can lead to coherence lengths smaller than the sizes of the crystallographic domains. In other words, lattice coherence lengths can be smaller than domain sizes seen by morphological probes such as atomic force microscopy.

\begin{figure}[ht]
  \includegraphics[width=6.2in]{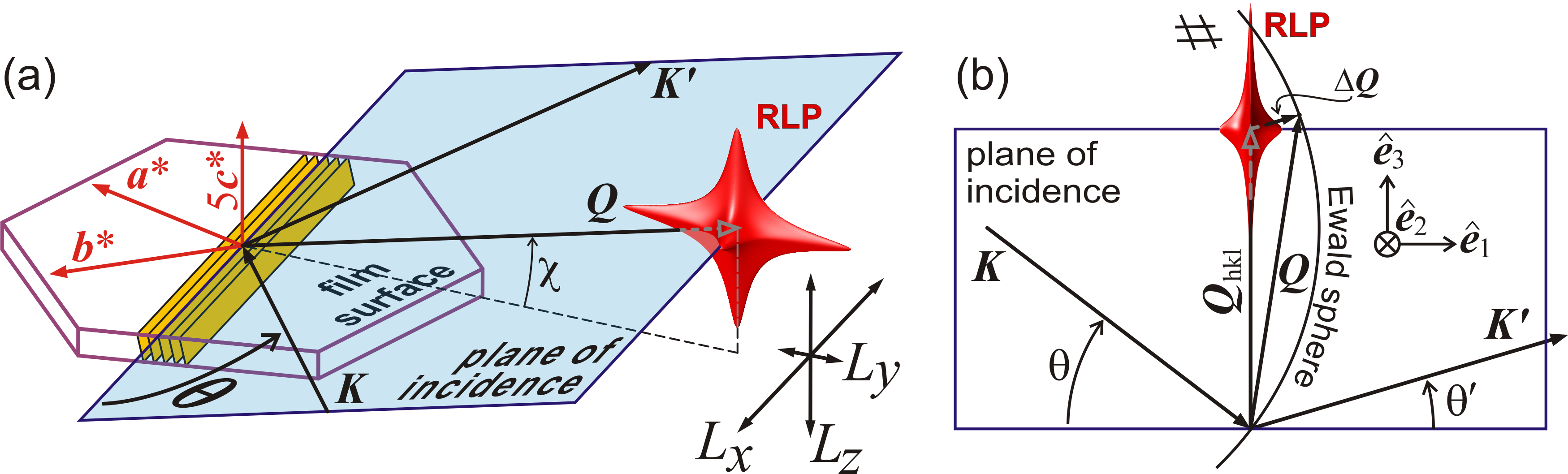}\\
  \caption{(a) Diffraction geometry of asymmetric reflection with diffraction vector $\vec{Q}=\vec{K}^\prime-\vec{K}$ at an angle $\chi$ from the film surface. Broadening of reciprocal lattice point (RLP) due to finite lattice coherence lengths $L_x$ and $L_y$ (along film in-plane directions), and $L_z$ (along the film growth direction). Hexagonal film with basis vectors $\vec{a}^*$, $\vec{b}^*$, and $\vec{c}^*$ of the reciprocal lattice. (b) Vectors, angles, and frame definitions used for calculating peak widths in rocking curve measurements ($\theta$-scans) around the reciprocal lattice vector $\vec{Q}_{\rm hkl} = {\rm h}\vec{a}^* +  {\rm k}\vec{b}^* +  {\rm l}\vec{c}^*$ of reflection hkl.  }\label{fig:qview}
\end{figure}

For diffraction vectors $\vec{Q}$ of asymmetric reflections, the film coherence lengths $L_x$, $L_y$, and $L_z$ are related to RLP broadening along in-plane directions
\begin{equation}\label{eq:xyinplane}
  \hat{\vec{x}} = \hat{\vec{z}}\times\vec{Q}/|\hat{\vec{z}}\times\vec{Q}|
  \quad {\rm and} \quad
  \hat{\vec{y}} = \hat{\vec{z}}\times\hat{\vec{x}}\,,
\end{equation}
as well as along the growth direction $\hat{\vec{z}}$, respectively. In diffraction geometry for very asymmetric reflections, as in Fig.~\ref{fig:qview}(a) where $\chi\ll90^\circ$, peak widths in rocking curve measurements ($\theta$-scans) are most susceptible to the RLP broadening along in-plane directions, Eq.~(\ref{eq:xyinplane}), since along the crystal truncation rod, \emph{i.e.} along $\hat{\vec{z}}$, it is nearly perpendicular to the incidence plane.

Lattice imperfections with Burgers vectors parallel to the Bragg plane are invisible to the corresponding Bragg reflection \cite{sm00}, producing no reduction of coherence lengths perpendicular to the diffraction vector, as in the case of $L_x$ for which $\vec{Q}\cdot\hat{\vec{x}}=0$. On the other hand, misfit of film/substrate parameters reduces the coherence length $L_y$ according to the average lattice imperfection separation distance \cite{jm78,mh04}
\begin{equation}\label{eq:latimperfdist}
    \bar{s} = \frac{a_s}{\varepsilon_0 + |\Delta a/a|}
\end{equation}
where $\varepsilon_0$ is the amount of misfit strain elastically accommodated in the absence of lattice mismatch $\Delta a/a = (a_f - a_s)/a_s$, regarding the actual film and substrate lateral lattice parameters present in the sample, $a_f$ and $a_s$ respectively.

In rocking curve measurements, diffraction peak widths are determined by the convolution between Ewald sphere and RLP as a function of the rocking angle $\theta$. If function $|W(\Delta\vec{Q})|^2$ describes the RLP broadening in reciprocal space, line profile of diffraction peaks can be calculated as \cite{sm16}
\begin{equation}\label{eq:Itheta}
    I(\theta)=\iint|W(\Delta\vec{Q})|^2\sin\theta^\prime{\rm d}\theta^\prime{\rm d}\varphi^\prime
\end{equation}
where $\Delta\vec{Q}=\vec{Q}-\vec{Q}_{\rm hkl}=[\vec{K}^\prime(\theta^\prime,\varphi^\prime)-\vec{K}(\theta)]-\vec{Q}_{\rm hkl}$ is the distance from the centre of the RLP given by the reciprocal lattice vector $\vec{Q}_{\rm hkl}$ of reflection hkl. For an incident wavevector written as $$\vec{K}=(2\pi/\lambda)[\cos\theta\,\hat{\vec{e}}_1 - \sin\theta\,\hat{\vec{e}}_3]$$ in the reference frame [$\hat{\vec{e}}_1,\,\hat{\vec{e}}_2,\,\hat{\vec{e}}_3$] of the incidence plane, as defined in Fig.~\ref{fig:qview}(b) where $\vec{Q}_{\rm hkl}=Q_{\rm hkl}\hat{\vec{e}}_3$, all physically possible wavevectors of diffracted x-rays (elastic scattering process) are accounted for as $$\vec{K}^\prime=(2\pi/\lambda)[\cos\theta^\prime\,\hat{\vec{e}}_1+ \sin\theta^\prime\sin\varphi^\prime\,\hat{\vec{e}}_2+ \sin\theta^\prime\cos\varphi^\prime\,\hat{\vec{e}}_3]\,,$$
even those rays going out of the incidence plane for which angle $\varphi^\prime\neq0$. Projection of $\Delta\vec{Q}$ in the $xyz$ frame earlier defined in Eq.~(\ref{eq:xyinplane}) is provided by
\begin{equation}\label{eq:M}
\hat{\vec{e}}_1 = \hat{\vec{x}},\quad
\hat{\vec{e}}_2 = \sin\chi\,\hat{\vec{y}} + \cos\chi\,\hat{\vec{z}},\quad{\rm and}\quad
\hat{\vec{e}}_3 = -\cos\chi\,\hat{\vec{y}} + \sin\chi\,\hat{\vec{z}},
\end{equation}
allowing the RLP broadening due to finite lattice coherence lengths along $\Delta Q_x=\Delta\vec{Q}\cdot\hat{\vec{x}}$, $\Delta Q_y=\Delta\vec{Q}\cdot\hat{\vec{y}}$, and $\Delta Q_z=\Delta\vec{Q}\cdot\hat{\vec{z}}$ to be taken into account for asymmetric reflections with diffraction vector at an angle $\chi$ from the film surface, Fig.~\ref{fig:qview}(a).

\begin{figure}
  \includegraphics[width=3.2in]{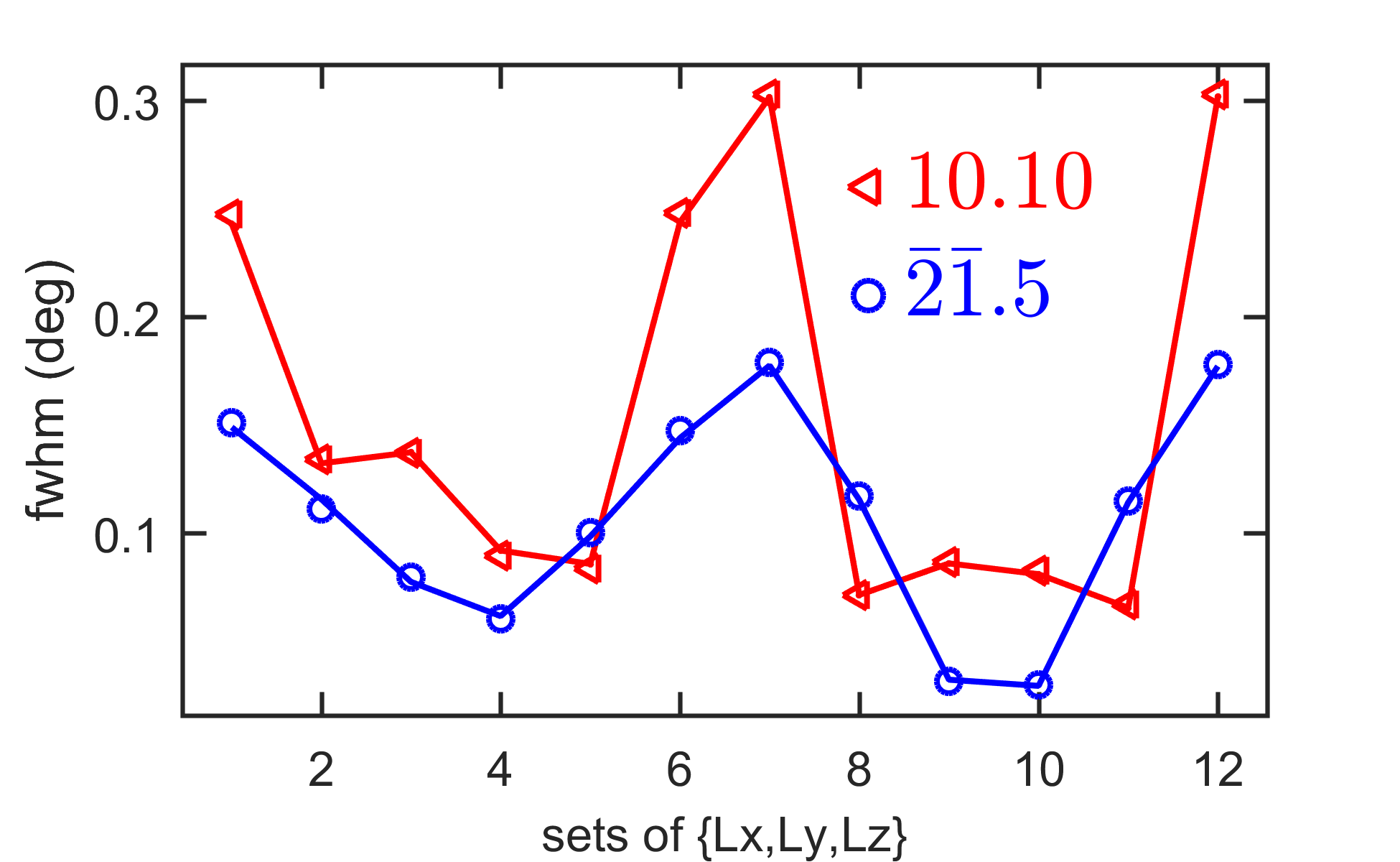}\\
  \caption{Comparison of peak widths (fwhm) from exact solution (symbols) in Eq.~(\ref{eq:Itheta}) and the proposed approach (solid lines) in Eq.~(\ref{eq:pwidth}). The $10\,10$ and $\bar{2}\bar{1}\,5$ reflections stand for the least and the most asymmetric ones listed in Table~\ref{tab:reflections}. \{$L_x$,$L_y$,$L_z$\} (nm) =  \{50,100,200\}, \{100,50,200\}, \{100,200,50\}, \{200,100,50\}, \{200,50,100\}, \{50,200,100\}, \{40,400,2000\}, \{400,40,2000\}, \{400,2000,40\}, \{2000,400,40\}, \{2000,40,400\}, and \{40,2000,400\} were the sets of coherence lengths used for this test, $1,2,\ldots,12$, respectively.
}\label{fig:rcfwhm}
\end{figure}

In one dimension, the Fourier transform of a finite lattice of length $L$ is the sinc function $\sin(\Delta Q\,L/2)/(\Delta Q/2)$ \cite{sm16}. Then, the modulus square of the normalized function
\begin{equation}\label{eq:WQ1}
W(\Delta\vec{Q})=
\frac{\sin(\Delta Q_x L_x/2)}{\Delta Q_x L_x/2}\,
\frac{\sin(\Delta Q_y L_y/2)}{\Delta Q_y L_y/2}\,
\frac{\sin(\Delta Q_z L_z/2)}{\Delta Q_z L_z/2}.
\end{equation}
has been chosen to describe the intensity distribution around the RLPs in Eq.~(\ref{eq:Itheta}). Although it is possible to fit experimental peak widths by handling numerically the double integral in Eq.~(\ref{eq:Itheta}), determination of the coherence lengths $L_{x,y,z}$ with this procedure can be very time consuming. Here, a different approach has been developed. Squared sinc functions have full width at half maximum (fwhm) given by $\beta_{x,y,z}=5.566/L_{x,y,z}$  (numerator comes from $\sin^2(x)/x^2=1/2$ when $4x=5.566$), which were projected in the incidence plane, and the corresponding peak widths  $\Delta\theta_{x,y,z}$ in $\theta$-scans obtained by using standard 2D Ewald construction in reciprocal space, \emph{e.g.} Fig.~\ref{fig:qview}(b). The resulting peak width is then calculated as
\begin{equation}\label{eq:pwidth}
    w_S = \sqrt{\Delta\theta_{x}^2 + \Delta\theta_{y}^2 + \Delta\theta_{z}^2}
\end{equation}
where $\Delta\theta_x = \beta_x/Q_{\rm hkl}$, $\Delta\theta_y = \beta_y\tan\theta\cos\chi/Q_{\rm hkl}$, and
$\Delta\theta_z = \beta_z\tan\theta\sin\chi/Q_{\rm hkl}$. In Fig.~\ref{fig:rcfwhm} there is a comparison of peak widths calculated by the exact solution in Eq.~(\ref{eq:Itheta}) and by the approach in Eq.~(\ref{eq:pwidth}). Since the latter approach shows very good agreement with the exact solution and is much faster in terms of CPU time, it has been used to determine the coherence lengths from the experimental peak widths.

Coherence length values were adjusted by SA algorithm to minimize the mean square deviation function $\sigma^2=\sum_j(w_E-w_S)_j^2/N_j$ where $w_E$ and $w_S$ stand for experimental and calculated peak widths, respectively. $w_S$ is obtained from Eq.~(\ref{eq:pwidth}). Subscript $j$ runs over the $N_j=3$ reflections in either hk\,5 and hk\,10 subsets of reflections. Uncertainties $\pm\sigma_L$ were estimated from the error bars $\sigma_w$ in $w_E$ values as $(\sigma_L/L)^2 = \sum_j(\sigma_w/w_E)_j^2$. The standard errors $\sigma_w$ were obtained by measuring a few times equivalent reflections set apart by 120$^\circ$ in azimuth (Table~\ref{tab:reflections}).

\section{S5 - Hybrid reflections}

Hybrid reflections have been studied and applied to investigate heteroepitaxial systems since 1981 \cite{ish81,sm91,sm93a,sm93b,sm98,sm03,sm07,men09,men10,jd16,es17,ep17}. However, only recently their occurance in epitaxial systems of hexagonal (001) films on cubic (111) substrates, such as Bi$_2$Te$_3$/BaF$_2$, have been predicted and observed at scattering angles \cite{sm18}
\begin{equation}\label{eq:tthnm}
    2\theta_{n,m} = 2\arcsin\left[\frac{\lambda}{2} \left(\frac{n}{a\sqrt{3}}+\frac{m}{c}\right)\right]
\end{equation}
where $n=\sum_s({\rm h}_s+{\rm k}_s+{\rm l}_s)>0$ and $m=\sum_f{\rm l}_f$. For the pair of hybrid reflections recently measured,\cite{sm19} hybrids $\bar{2}2.\bar{10}_f+044_s$ (peak f/s) and $404_s+0\bar{2}.\bar{10}_f$ (peak s/f), both have $n=8$ and $m=-10$. By using $a=6.2001$\,\AA\, as the cubic lattice parameter of BaF$_2$ and $c=30.497$\,\AA\, as the hexagonal lattice parameter of the film, $\theta_{8,-10}=18.74^\circ$ is close to the incidence angle used to excite these hybrids in symmetric diffraction geometry. However, each hybrid occurs at different azimuth. For the reference of azimuth defined in Fig.~\ref{fig:recspaceviews}, peak f/s is centred at about $\varphi=53.6^\circ$ and peak s/f at $\varphi=66.4^\circ$. Meshscans in $\theta$ and $\varphi$ were carried out around these azimuths to proper determine the hybrid peak position in $\theta$; a detailed description on how to measure such hybrids can be found elsewhere \cite{sm18}. The split of a hybrid pair as function of the rocking curve angle $\theta$ is proportional to $\Delta a/a$ as given by  \begin{equation}\label{eq:thincxphi}
    \Delta\theta \simeq -2\frac{\vec{Q}_{f,\|}\cdot\hat{\vec{k}_{\|}}}{Q^*}\frac{\Delta a}{a}\,.
\end{equation}
$\vec{Q}_{f,\|}$ is the in-plane component of the film diffraction vector and $\hat{\vec{k}_{\|}}$ is the in-plane direction of the incident wavevector. For the case of hybrids $\bar{2}2.\bar{10}_f+044_s$ (peak f/s) and $404_s+0\bar{2}.\bar{10}_f$ (peak s/f),   $\vec{Q}_{f,\|}\cdot\hat{\vec{k}_{\|}}/Q^*=1.0176$, which leads to the values of $\Delta a/a$ reported here.

\section{S6 - Film composition}

\begin{figure}
  \includegraphics[width=6.4in]{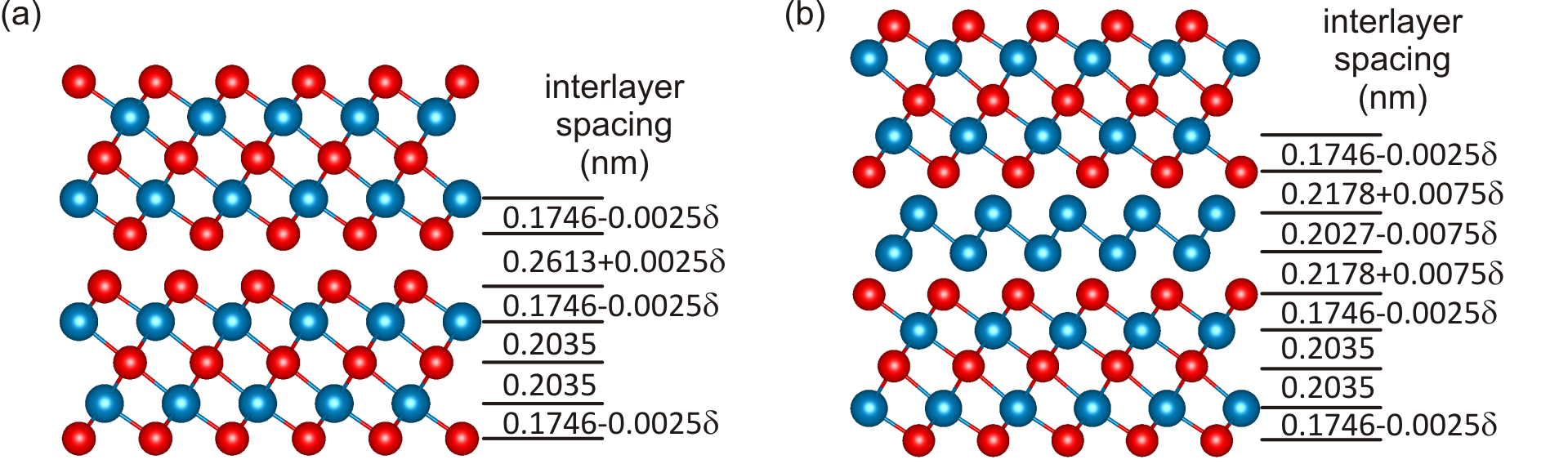}\\
  \caption{Interlayer spacing for model structures as a function of Te deficit $\delta$ in films with composition Bi$_2$Te$_{3-\delta}$. (a) Around the vdW gap between adjacent QLs, and (b) when a bismuth bilayer (BL) is formed in the vdW gap. First order approximation of structural strain caused by the presence of BLs (see Appendix of  Ref. 29 for more details).}
  \label{fig:interspacing}
\end{figure}

\begin{figure}
  \includegraphics[width=6.4in]{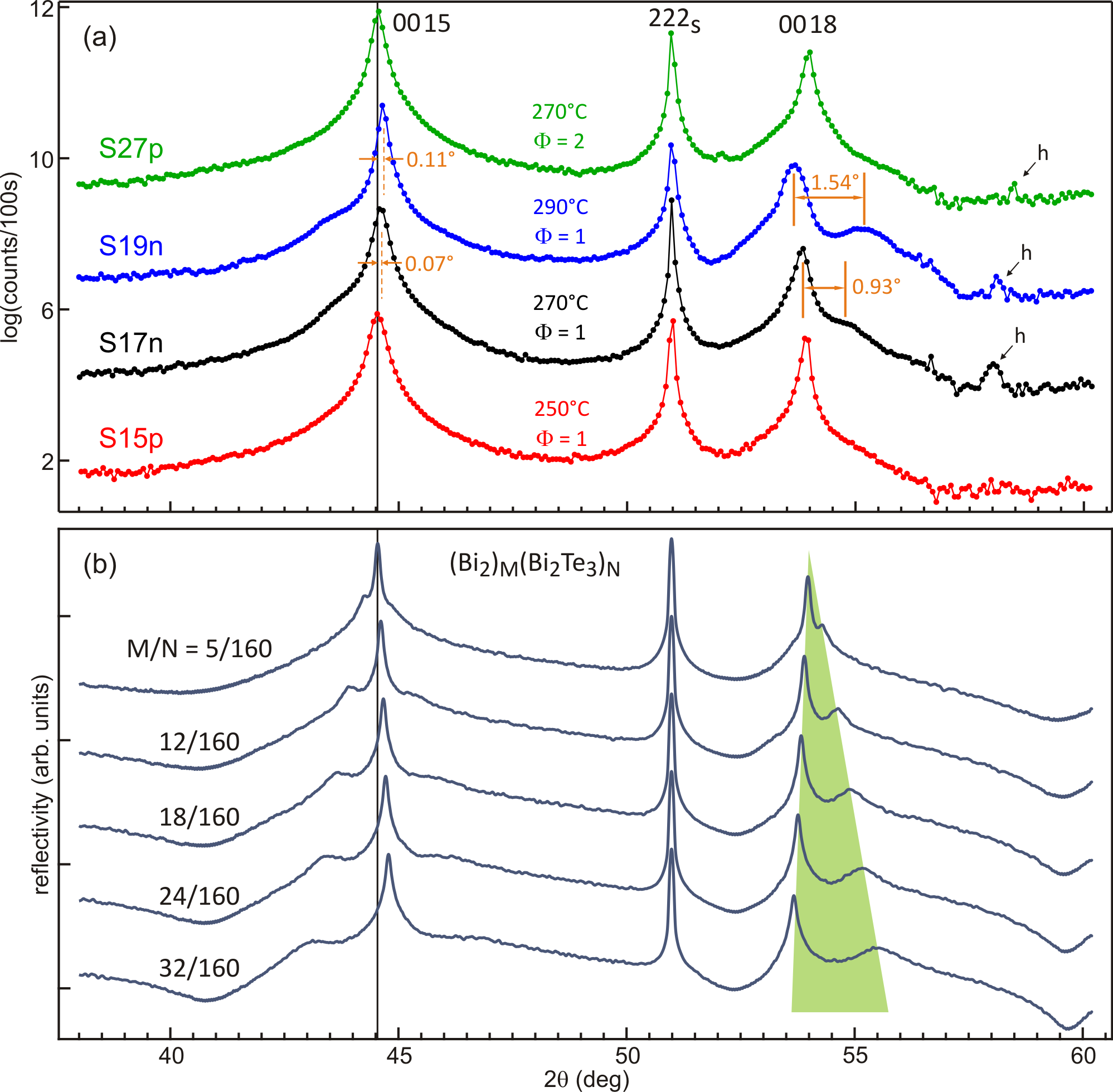}\\
  \caption{(a) $\theta/2\theta$-scans along the $c$-axis of Bi$_2$Te$_3$ films on BaF$_2$ (111) substrates. X-rays of $\lambda=1.540562$\,\AA\, (Cu$K_{\alpha1}$). Sample labels, substrate temperature during film growth, and ratio $\Phi$ between beam equivalent pressures of Te and Bi$_2$Te$_3$ sources are indicated at each curve. Small peaks of hybrid reflections are seen (arrows h).\cite{sm18} Shifting of $00\,15$ film reflection towards the $222_s$ substrate reflection, as well as splitting of $00\,18$ reflection peak, are due to the presence of BLs in the film structure. (b) Simulation of $\theta/2\theta$-scans for $({\rm Bi}_2)_M({\rm Bi}_2{\rm Te}_3)_N$ films with number $M$ of BLs per number $N$ of QLs by using a recursive series for x-ray dynamical diffraction calculation, as introduced elsewhere \cite{sm17}. The splitting of peak $00\,18$ (shaded area) is proportional to the content of BLs or, equivalently, to the Te deficit $\delta$ in Bi$_2$Te$_{3-\delta}$. It leads to 15/160 ($\delta \simeq 0.26$) and 26/160 ($\delta \simeq 0.42$) as the relative number $M/N$ of BLs in the samples S17n and S19n, respectively.}\label{fig:Lscans}
\end{figure}

Films with composition $({\rm Bi}_2)_M ({\rm Bi}_2{\rm Te}_3)_N$ have deficit $\delta=3M/(N+M)$ of tellurium due to formation of bismuth bilayers (BLs) in the vdW gap between adjacent quintuple layers (QLs). X-ray diffraction simulation in model structures containing a number $M$ of BLs randomly distributed along the film thickness have been used to determine the actual composition of the films.\cite{hs14,sm17,gs18} In first order approximation, variation of interlayer spacing in the model structures as a function of $\delta$ were accounted for as shown in Fig.~\ref{fig:interspacing}. By comparing experimental and simulated x-ray diffraction curves in Fig.~\ref{fig:Lscans}, only two samples present features owing to the presence of BLs: shifting of peak $00\,15$ whose position is determined by the mean interlayer spacing $\langle d \rangle \simeq 0.2035-0.0025\delta$;\cite{hs14,cf16a} and splitting of peak $00\,18$ (shaded area in Fig.~\ref{fig:Lscans}(b)) that is also proportional to $\delta$ according to $(2\pi/\lambda)\cos\theta\Delta2\theta=0.23\,\delta$ (for $\Delta2\theta$ in radians) \cite{cf16a,sm17}. By using this later formula with the values indicated in Fig.~\ref{fig:Lscans}(a), samples S17n and S19n have films of compositions Bi$_2$Te$_{2.74}$ and Bi$_2$Te$_{2.58}$, respectively.

\bibliography{dynamicdefectBi2Te3suppinfo}

\end{document}